\documentclass[twocolumn,amsmath,floatfix,prl,aps]{revtex4-1}

\usepackage{color}
\usepackage{lipsum,amsmath}
\usepackage{pifont}   % Ding symbols
\usepackage{graphicx} % Include figure files
\usepackage{dcolumn}  % Align table columns on decimal point
\usepackage{bm}       % bold math
\usepackage{amsfonts} % Some more fonts
\usepackage{amssymb}  % More symbols
\usepackage{multirow} % Table functions
\usepackage{tikz}
\usepackage{siunitx}
\usepackage{placeins}
\usepackage{braket}
\usepackage{nicefrac}
\usepackage{bm}
\usepackage{physics}
\usepackage{grffile}

\usepackage{array,multirow,graphicx}
\usepackage{ulem}
\usepackage[latin1]{inputenc}
\usepackage{caption}
\usepackage{scrextend}
\usepackage{hyperref}
\usepackage{float}
\usepackage{subcaption}
\usepackage{gensymb}

\newcommand{\ben}{\begin{equation}}
\newcommand{\een}{\end{equation}}
\def\bk{{\bf k}}

\usepackage{amsmath}
\usepackage{amsfonts}
\usepackage{amssymb}
\usepackage{graphicx}
\usepackage{siunitx}

\begin{document}

\title{Making a case for femto- phono- magnetism with FePt}

\author{S. Sharma}
\affiliation{Max-Born-Institut  f\"ur Nichtlineare Optik und Kurzzeitspektroskopie, Max-Born-Strasse 2A, 12489 Berlin, Germany.}
\author{S. Shallcross}
\affiliation{Max-Born-Institut  f\"ur Nichtlineare Optik und Kurzzeitspektroskopie, Max-Born-Strasse 2A, 12489 Berlin, Germany.}
\author{P. Elliott}
\affiliation{Max-Born-Institut  f\"ur Nichtlineare Optik und Kurzzeitspektroskopie, Max-Born-Strasse 2A, 12489 Berlin, Germany.}
\author{J.~K. Dewhurst}
\affiliation{Max-Planck-Institut f\"ur Mikrostrukturphysik, Weinberg 2, D-06120 Halle, Germany.}
\email{sharma@mbi-berlin.de}

\date{\today}

\begin{abstract}
In the field of femtomagnetism magnetic matter is controlled by ultrafast laser pulses; here we show that coupling phonon excitations of the nuclei to spin and charge leads to femto-phono-magnetism, a powerful route to control magnetic order at ultrafast times. With state-of-the-art theoretical simulations of coupled spin- charge- and lattice-dynamics we identify strong non-adiabatic spin-phonon coupled modes that dominate early time spin dynamics. Activating these phonon modes we show leads to an additional (up to 25\%) loss of moment in FePt occurring within 40 femtoseconds of the pump laser pulse. Underpinning this enhanced ultrafast loss of spin moment we identify a physical mechanism in which minority spin-current drives an enhanced inter-site minority charge transfer, in turn promoting increased on-site spin flips. Our finding demonstrates that the nuclear system, often assumed to play only the role of an energy sink aiding long time re-magnetisation of the spin system, can play a profound role in controlling femtosecond spin-dynamics in materials.
\end{abstract}

\maketitle

\section*{Introduction}

The speed at which information is stored and processed is determined by the fundamental time scales at which  matter can be manipulated by external fields\cite{basov}. The fastest such control is through the interaction of matter with the electromagnetic field of light, with ultrafast laser control over magnetic order now well established as a key low energy route to controlling microscopic order\cite{kirilyuk10}. To leading order, however, the electromagnetic field does not couple directly to spin\cite{itoh}, and so control over magnetic order proceeds indirectly through charge excitations. 

At femtosecond time scales this can occur either through spin transfer between magnetic sub-lattices\cite{dewhurst2018,Siegrist2019}, through manipulation of spin-orbit coupling\cite{krieger2015,ZH00,substrate18}, or at later time scales via the exchange interaction\cite{exchange}. In all these processes the lattice merely plays the role of an energy and momentum reservoir, either for absorbing the momentum of demagnetized matter or, at later time scales, remagnetising matter through equilibration\cite{baum22,dornes2019}. 

Recently, the lattice has been employed as a route to directly control matter through the excitation of infra-red phonon modes, representing another route by which matter can be controlled by light\cite{rini07,man17,nova19,bargheer21,mash21,marg}. The idea of phono-magnetism\cite{afa21,narang} i.e. manipulation of spins via nuclear dynamics is only now beginning to be examined as an additional degree of freedom via which spin can be dynamically manipulated. While it has been shown to represent a viable route to control magnetic order at longer picosecond time scales, the role that the lattice can play in the manipulation of magnetic matter at the fastest possible femtosecond and sub-femtosecond time scales remains unexplored and this is the central question we will address in this work.

As far as theoretical treatment of light-matter interaction is concerned, time-dependent density functional theory (TDDFT)\cite{RG1984} is a fully quantum mechanical and {\it ab-initio} method that has been shown to be very successful not just in providing insight into experiments but also in predicting new phenomena\cite{dewhurst2018,Siegrist2019,hofherr2020,clemens2020}. In the present work we provide a method, suitable for extended solids, via which charge and spin dynamics within TDDFT can be coupled to the motion of the nuclei. Using this sophisticated technique and with the example of FePt we address the question of what role lattice phonon excitations play in the dynamics of magnetic order on ultrafast time scales. Our key finding is that selective lattice excitation can result in substantially enhanced demagnetisation, behaviour in striking contrast to the commonly held view that the lattice aids re-magnetisation at longer time scales. Furthermore, we find that this effect is both profoundly non-adiabatic as well as non-linear in phonon amplitude.

\section*{Results}

\begin{figure}[ht]
\includegraphics[width=\columnwidth, clip]{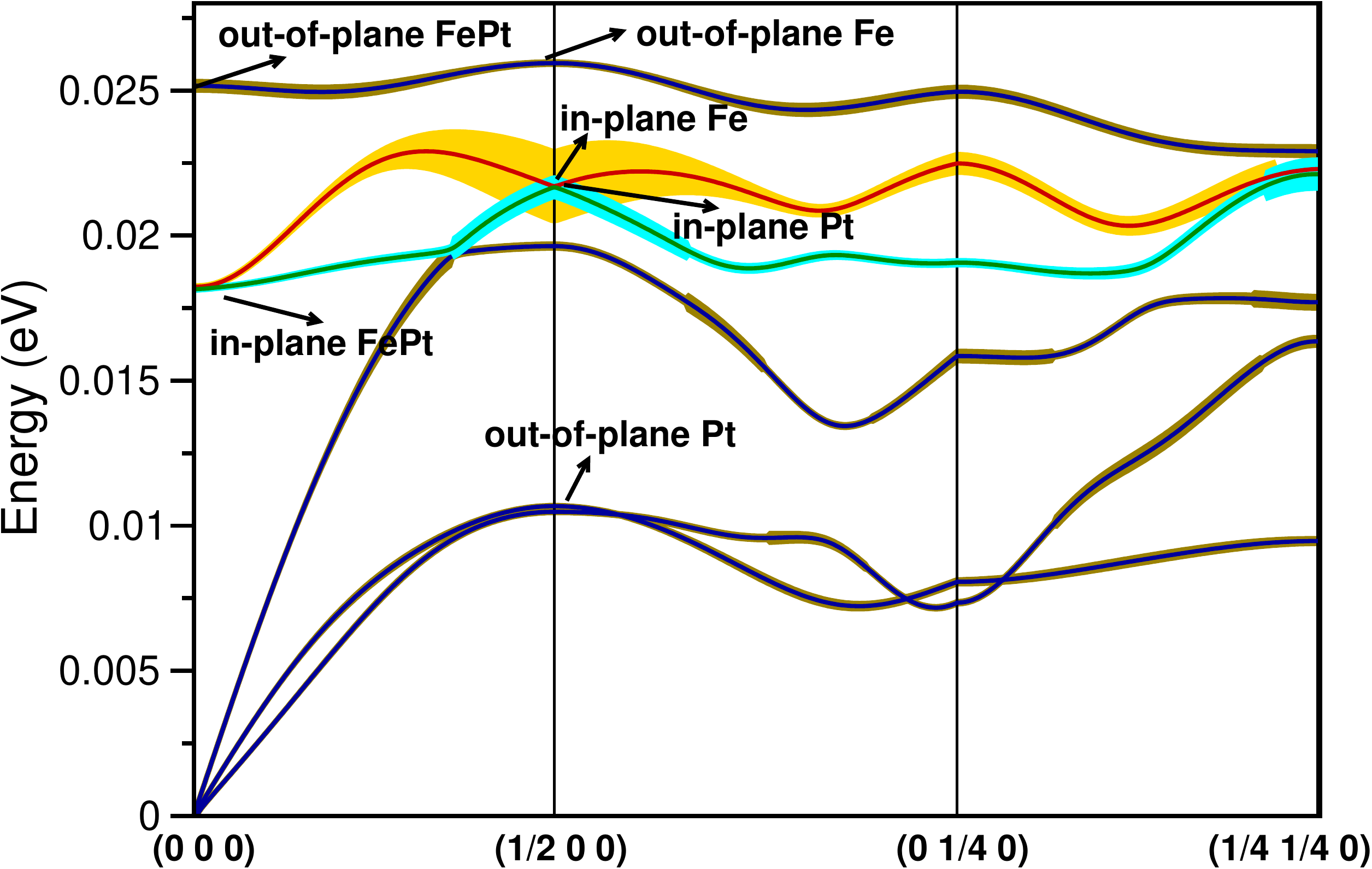}
\caption{\raggedright \raggedright Phonon spectra for FePt with electron-phonon coupling strength indicated by the width of the spectral line. As can be seen, electron-phonon coupling is strongest at the X-point (for details of the nature of these modes see the legend). Here in-plane refers to the Fe and Pt planes of the L1$_0$ structure, stacked perpendicular to the $c$-axis that is also the direction of ground state magnetization.} \label{phon}
\end{figure}

\paragraph*{Electron phonon coupling} In order to manipulate spin-dynamics via selected phonon modes as a first step we examine the phonon spectra and electron-phonon coupling (EPC), shown in Fig.~\ref{phon} for FePt. The choice of the material is based on the fact that it is known to have strong coupling of spins with lattice\cite{reppert20}.  Evidently, the EPC is strongest for the optical phonon modes at the X ($1/2,0,0$) point\cite{maldo17}, and an analysis of the phonon modes indicates that the two strongly coupled modes are the pure in-plane Fe mode and the pure in-plane Pt mode with a period $\sim$190~fs. With this information in hand we now study the dynamics of the the pump laser excited spin system in presence of such modes. In doing so we have in mind a setup where these phonons are pre-excited, following which an optical laser pump pulse is used to manipulate spins in presence of the excited phonon modes. 

\begin{figure}[ht]
\includegraphics[width=\columnwidth, clip]{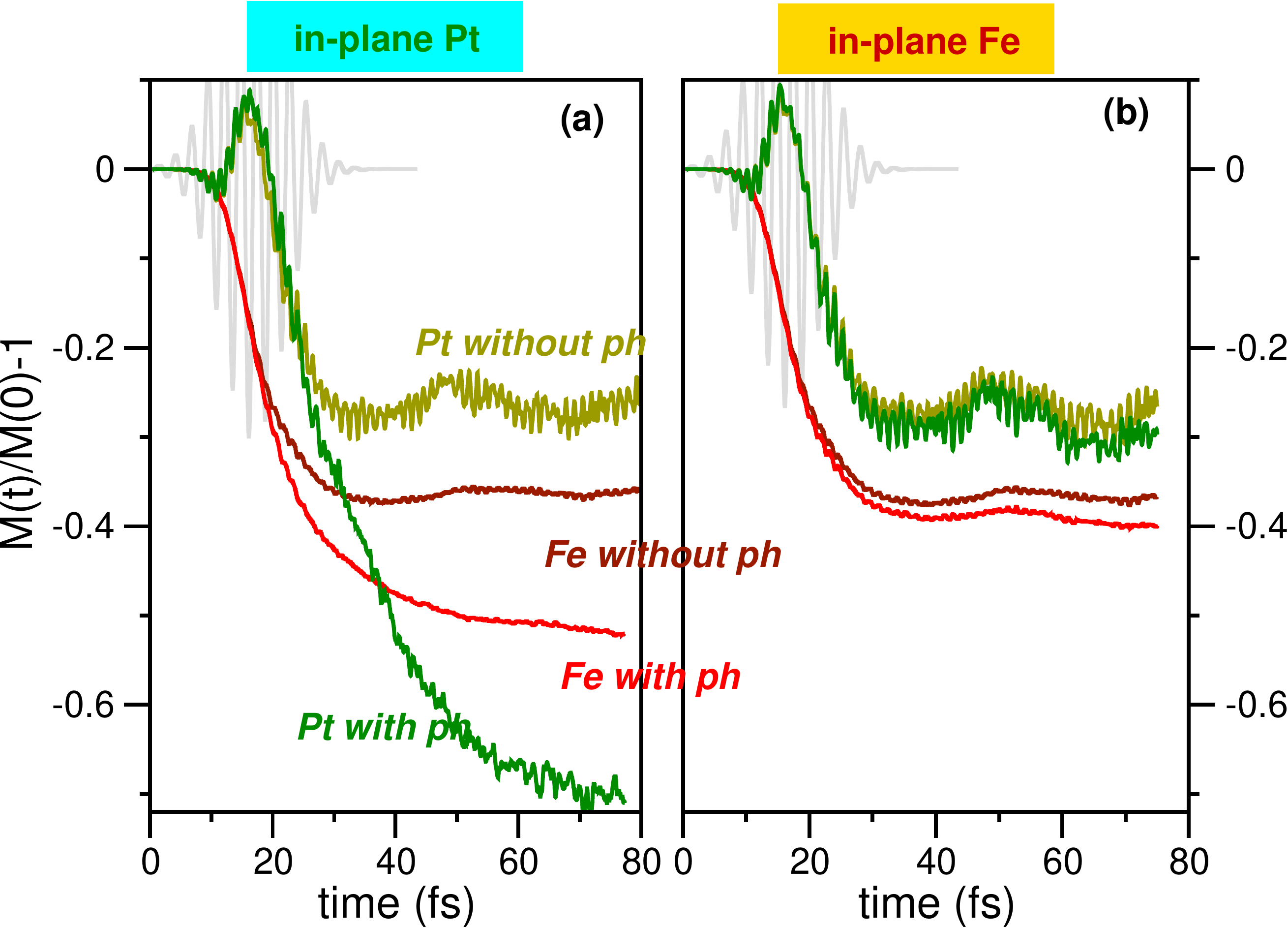}
\caption{\raggedright \raggedright Normalized atom-resolved spin moment as a function of time (in fs) in laser pumped FePt. The pump pulse (the vector potential is shown in grey) has a central frequency of 1.55~eV, incident fluence 9~mJ/cm$^2$ (i.e. amplitude 2x10~$^{12}$W/cm), FWHM of 13~fs, and is polarized in the plane of the phonon mode. Spin dynamics calculations are performed both in the presence and absence of a phonon excitation, with results shown for the two most strongly coupled phonon-modes at the X-point: (a) the in-plane Pt mode and (b) the in-plane Fe mode.} \label{mom}
\end{figure}

\paragraph*{Femto-phono-magnetism} In Fig.~\ref{mom} are shown the results for the spin-dynamics under the influence of a pump pulse and in presence and absence of the two strongly coupled X-point phonon modes, with the simulation time equal to approximately a half-cycle of these modes.  The time difference between the pulse envelope maxima and the phonon amplitude maxima (at 0fs) is in this work set to 20~fs. For these modes the two Fe (or Pt) atoms move in-plane (perpendicular to the spin quantization axis) and in mutually opposite directions.
Surprisingly we see that the in-plane Pt mode has a significant effect on the spin-dynamics with demagnetization of the Fe atom increasing by 0.5~$\mu_B$ per Fe atom (20\% of the ground state moment), an effect that occurs in less that 40~fs after the pump pulse (see Fig.~\ref{mom}). At early times ($<25$~fs) the moment on Fe lattice decreases with a corresponding increase in the moment of the Pt sub-lattice, a signature of OISTR (optical inter-site spin transfer)\cite{dewhurst2018} (see Fig.~S2 of SI). The induced moment on Pt site subsequently decreases and also shows a profoundly different spin-dynamics in the presence and absence of nuclear dynamics. These results highlight the fact that the nuclear dynamics, long assumed to be important only as an energy and momentum reservoir, can in fact, when selectively pre-excited, decisively influence ultrafast demagnetization allowing for faster spin-manipulation times.

\begin{figure}[ht]
\includegraphics[width=\columnwidth, clip]{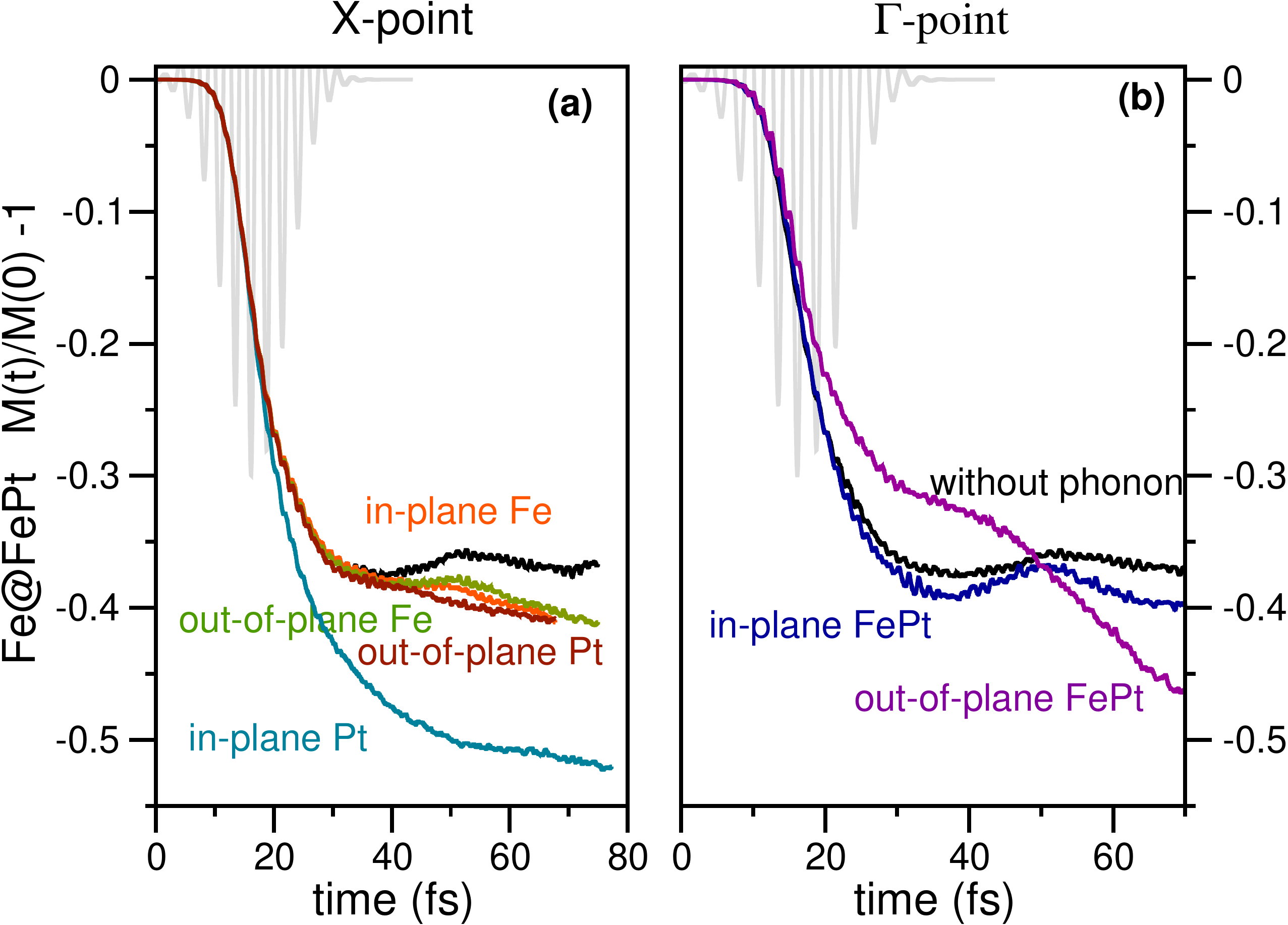}
\caption{\raggedright \raggedright Normalized Fe atom projected spin moment as a function of time (in fs) in laser pumped FePt. The same pump pulse (vector potential is shown in grey) as in Fig.~\ref{mom} is used. The moment is calculated in presence of phonon modes as well as in absence of any nuclear motion. Results are shown for four different phonon-modes at (a) the X-point and (b) $\Gamma$-point (see legend and Fig.~\ref{phon} for mode details about the mode). For a direct comparison the amplitude of all the modes is kept constant at 1.1~pm.} \label{mom-all}
\end{figure}

Even though the in-plane Fe mode has a very strong EPC the effect of this mode on the spin dynamics is insignificant at early times. This indicates that large EPC, while useful as a guide, does not necessarily imply large spin-phonon coupling; out of all the X and $\Gamma$ point phonon modes studied in the present work the in-plane Pt mode contributes most significantly to spin-manipulation (see Fig.~\ref{mom-all}).

The formalism used for coupling nuclear dynamics to electronic degrees of freedom is exact for small displacements of the nuclei and, for the results discussed thus far, the phonon amplitudes are set to 1.1~pm for all modes. Such small nuclear displacements by themselves have no significant effect on the spin magnetic moment in the ground-state (with the change in ground state moment due to this atomic displacement in a static calculation found to be 0.009~$\mu_B$ per Fe atom and 0.0008~$\mu_B$ per Pt atom). The effect of nuclear degrees of freedom on spins is therefore profoundly non-adiabatic. 

\begin{figure}[ht]
\includegraphics[width=\columnwidth, clip]{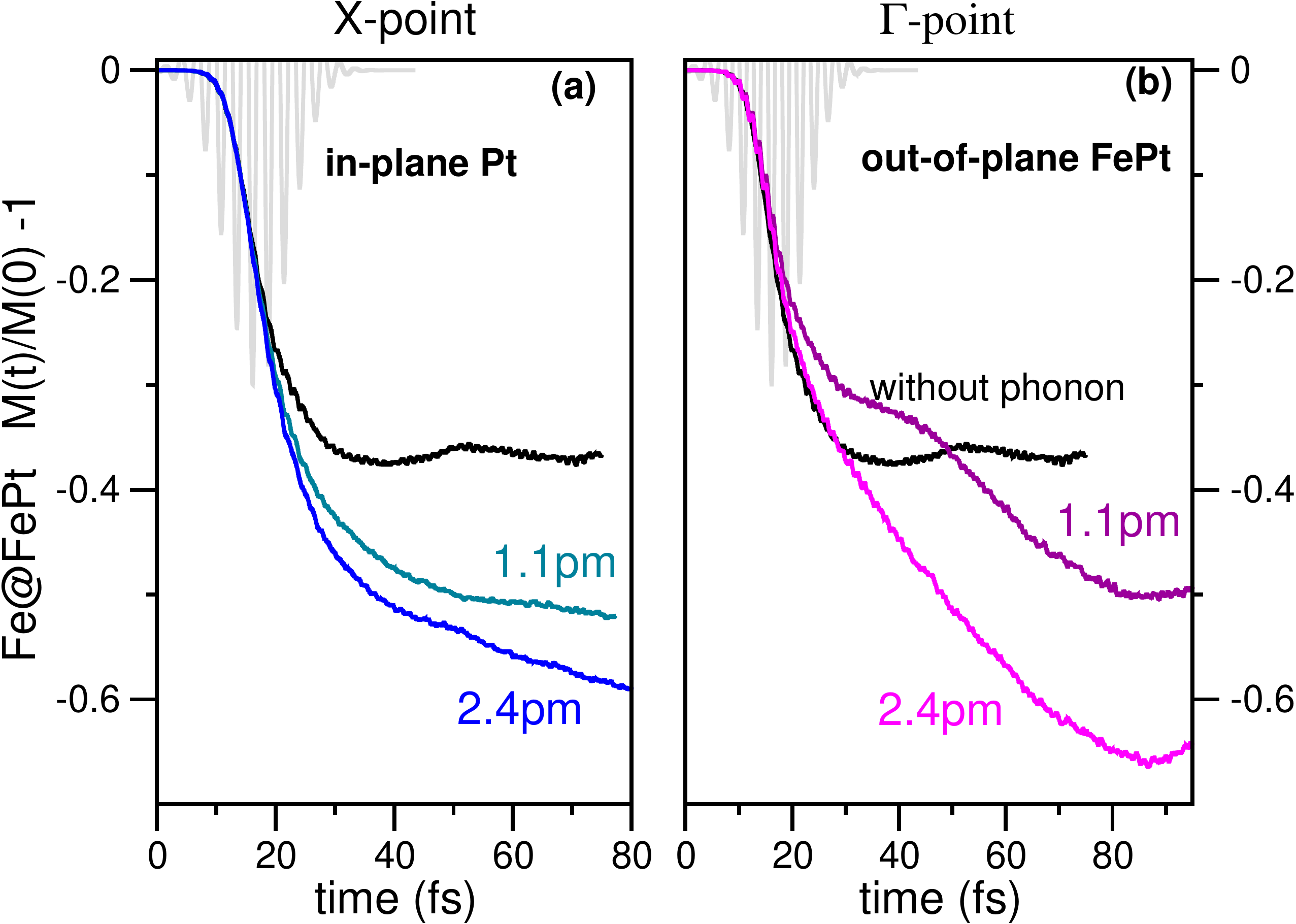}
\caption{\raggedright \raggedright Normalized Fe atom projected spin moment as a function of  time (in fs) in laser pumped FePt. The pump pulse (vector potential is shown in grey) has central frequency of 1.55~eV, incident fluence 9~mJ/cm$^2$ (i.e. amplitude 2x10~$^{12}$W/cm) and FWHM of 13~fs polarized in the plane of the phonon mode. The moment is calculated in presence of phonon modes as well as in absence of any nuclear motion. Results are shown phonon-modes with two different amplitudes at (a) the X-point and (b) $\Gamma$-point} \label{Lamp}
\end{figure}

\paragraph*{Infra red active phonons} From the results presented thus far it is clear that for small amplitude (1.1~pm) the X-point phonons dominate the spin dynamics. However selective excitation of such finite momentum phonons requires considerable effort\cite{abebe19,samn15}, with the experimentally easily accessible phonon excitations being the infra-red (IR) active phonon modes at $\Gamma$-point. A crucial question is therefore how such phonons impact the spin dynamics once the small amplitude restriction is relaxed. We now increase the phonon amplitudes to 2.4pm (see Fig.~\ref{Lamp}). It is important to note that these amplitudes are still small, well within the range what is experimentally naturally excited\cite{baum22}. Remarkably, for such amplitudes a new feature of femto-phono-magnetism emerges with the IR active (4.3~THz) out-of-plane Fe-Pt mode (in which the Fe and Pt planes oscillate parallel to the spin quantization axis and in opposite directions to each other) now significantly impacting the magnetization dynamics, with an extra 0.8$\mu_B$ (25\%) per Fe atom demagnetization at 65~fs. Reducing the pump laser pulse fluence to 2~mJ/cm$^2$ this enhancement of demagnetization increases to 40\% (see Fig.~S3 of the SI).  While the X-point phonon mode also shows further increase in demagnetization, it is no longer the most important mode for spin manipulation. These results highlight that the spin-nuclear coupling is not only highly non-adiabatic, but also has a strong non-linear dependence on phonon amplitude. That the IR active mode can also be used to strongly manipulate spins makes the present results highly significant and directly accessible by experiments.

\begin{figure}[ht]
\includegraphics[width=\columnwidth, clip]{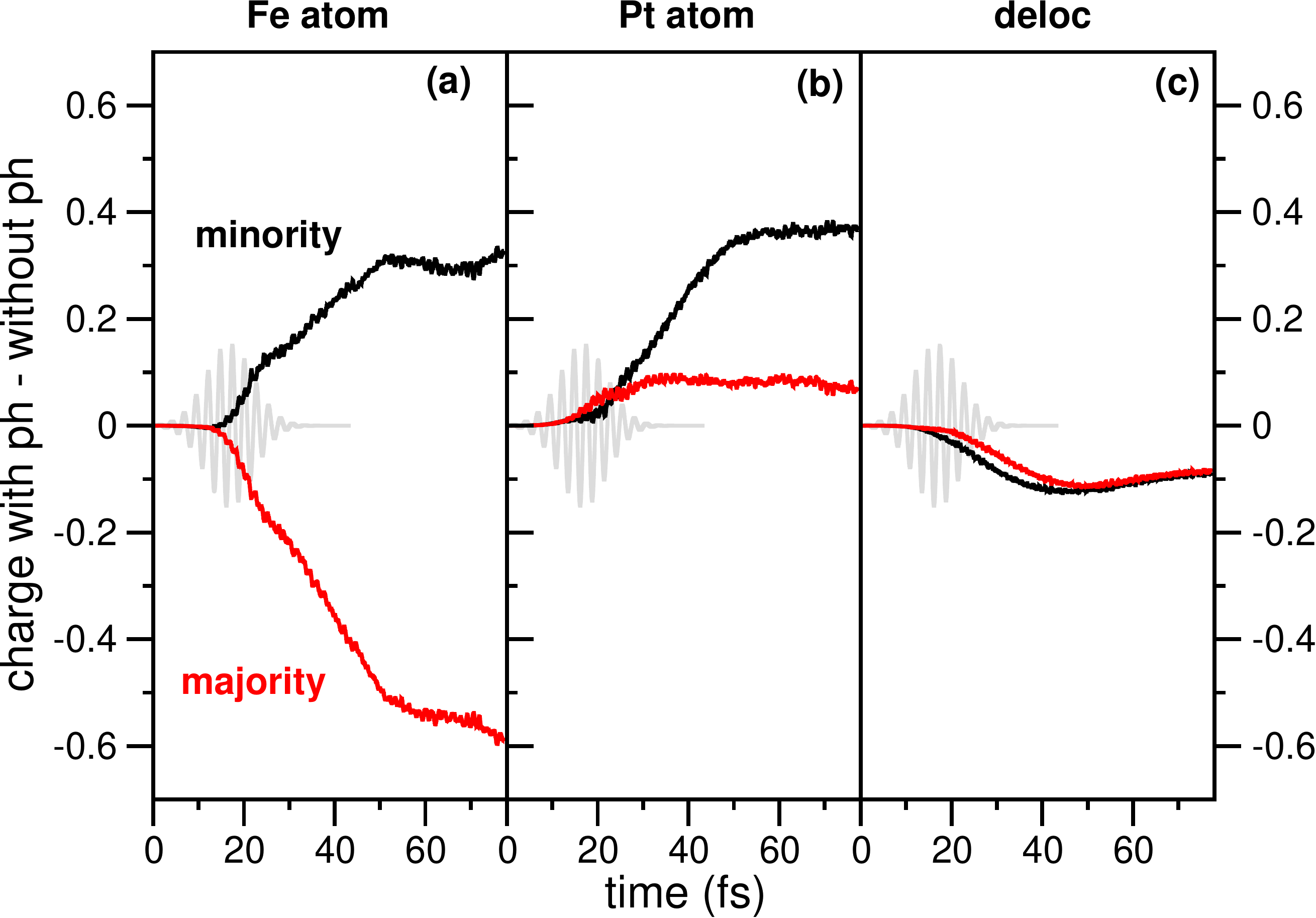}
\caption{\raggedright \raggedright Difference in atom and spin resolved charge calculated with and without phonons for laser pumped FePt. The calculations with phonons were performed in presence of the in-plane Pt mode. The same pump pulse (vector potential shown in grey in each panel) is used as in Fig.~\ref{mom}. Results are shown for (a) the Fe atom, (b) the Pt atom and (c) delocalized charge in highly excited states that cannot be assigned to either atom.} \label{chg}
\end{figure}

\paragraph*{Spin current}  In order to further probe the reasons behind this enhanced demagnetization due to non-adiabatic coupled spin-nuclear dynamics we now look at the transient spin dependent charge. In Fig.~\ref{chg} is plotted the difference between transient charge with and without inclusion of nuclear dynamics, shown projected onto spin and atom type. Both with and without phonons the same laser pump pulse of central frequency 1.55~eV, duration 15~fs, and incident fluence of 9~mJ/cm$^2$ (amplitude of 2$\times$10$^{12}$~W/cm$^2$) was used. As can be seen, in the coupled nuclear and spin dynamics the main
role is played by minority charge with the presence of a pre-excited phonon shown to result in a substantially increased flow of minority spin-current from Fe to Pt. This, in turn, dynamically increases the number of available Fe minority states leading to increased majority to minority spin-flips on the Fe site and a large demagnetization of both Fe and Pt sublattices. This flow of minority spin current from Fe to Pt atoms is due to large number of available Pt minority states (see the DOS for FePt in SI Fig.~S1). It has been recently shown that such currents can be very important for THz generation\cite{runge21}, which is a radiative effect and is of crucial importance at later times. In future it would be interesting to study light emission caused by phono-magenetism.
Another effect of nuclear-spin coupling we find is a localization of high-energy delocalized electrons, with delocalized charge decreasing in both spin channels (see Fig. ~\ref{chg}(c)). This physics has recently also been seen in graphite where high energy charge (6-8~eV above the Fermi level) is localized and stabilised due to phonons\cite{themis}. The highlight of these results thus is that the minority spin-current between sub-lattices controls the physics in the case of non-adiabatic and non-linear (in amplitude) spin-phonon coupling. 

\section*{Discussion}

We demonstrate that (a) nuclear degrees of freedom do not merely act as energy sink for pumped spins, but can also be used to enhance spin-dynamical effects at femtosecond time scales, (b) that this enhancement occurs through profoundly non-adiabatic spin-phonon coupling and, (c) that the microscopic mechanism underpinning this effect of femto-phono-magnetism is a substantial increase of the minority spin current flow between sub-lattices with concomitant increase SOC spin flips into empty minority states, in turn resulting in enhanced demagnetization.
The coupling between phonons and the spin dynamics on femtosecond time scales is found not only to be non-adiabatic, but also highly non-linear: at an amplitude of 1.1~ps the X point phonons dominate while doubling this amplitude (still small and experimentally feasible) leads to $\Gamma$-point (IR active) dominance of the spin-phonon coupling.

These results open a new route towards spin-manipulation by small amplitude coherent phonons in femtosecond spin dynamics. The mechanism of inter-site minority current flow suggests that multi-component magnetic materials will present a rich field for spin manipulation via nuclear degrees of freedom, with similar physics expected to be found at interfaces between magnetic materials or a magnetic and non-magnetic material, in which interface currents will be controlled by interface phonon modes. In future it would also be interesting to explore the effects of such currents on light emission.  Our results thus suggest a powerful new route to controlling magnetic order at femtosecond time scales with the nuclear and electron systems both now viewed as key degrees of freedom for the ultrafast manipulation of the spin degree of freedom.

\section*{Materials and Methods}

To study the dynamics of the spin and charge we have used  fully {\it ab-initio} state-of-the-art time dependent density functional theory (TDDFT), which rigorously maps the computationally intractable problem of interacting electrons to the Kohn-Sham (KS) system of non-interacting fermions in a fictitious potential, which can be solved by modern computing clusters. The time-dependent KS equation is:

\begin{align}
i &\frac{\partial \psi_{j\bk}({\bf r},t)}{\partial t} =
\Bigg[
\frac{1}{2}\left(-i{\nabla} +\frac{1}{c}{\bf A}_{\rm ext}(t)\right)^2 +v_{s}({\bf r},t) \nonumber \\
& + \frac{1}{2c} {\sigma}\cdot{\bf B}_{s}({\bf r},t) + \frac{1}{4c^2} {\sigma}\cdot ({\nabla}v_{s}({\bf r},t) \times -i{\nabla})\Bigg]
\psi_{j\bk}({\bf r},t)
\label{TDKS}
\end{align}
where ${\bf A}_{\rm ext}(t)$ is a vector potential representing the applied laser field. It is assumed that the wavelength of the applied laser is much greater than the size of a unit cell and the dipole approximation can be used i.e. the spatial dependence of the vector potential is disregarded. The KS fictitious potential $v_{s}({\bf r},t) = v_{\rm ext}({\bf r},t)+v_{\rm H}({\bf r},t)+v_{\rm xc}({\bf r},t)$ is decomposed into the external potential $v_{\rm ext}$, the classical electrostatic Hartree potential $v_{\rm H}$ and the exchange-correlation (XC) potential $v_{\rm xc}$. Similarly, the KS magnetic field is written as ${\bf B}_{s}({\bf r},t)={\bf B}_{\rm ext}(t)+{\bf B}_{\rm xc}({\bf r},t)$ where ${\bf B}_{\rm ext}(t)$ is an external magnetic field and ${\bf B}_{\rm xc}({\bf r},t)$ is the exchange-correlation (XC) magnetic field. ${\sigma}$ are the Pauli matrices. The final term of Eq.~\eqref{TDKS} is the spin-orbit coupling term.

The back-reaction of the nuclear dynamics on the electronic system is approximated with:

\begin{equation}
    v_{s}({\bf r},t) \longrightarrow v_{s}({\bf r},t)-\sum_{p \alpha} \frac{\partial v_{\rm cl}({\bf r})}{\partial u^p_{\alpha}} \delta u^p_{\alpha}(t),
\label{gradv}
\end{equation}
where $p$ labels a nucleus and $\alpha$ is the direction. $v_{\rm cl}$ is the Coulomb potential from the nucleus and the core density of electrons. $\delta u^p_{\alpha}(t)$ is the displacement of the nucleus away from equilibrium and is calculated from the atomic forces\cite{force}. There are two contributions to the forces generated on the nuclei that lead to nuclear dynamics: (a) motion of electronic charge due to the pump laser pulse \cite{shino10} and (b) the forces due to nuclei being out of equilibrium caused by excitation of phonon modes. These forces then cause the nuclei to move according to the classical equations of motion. These moving nuclei themselves cause a back-reaction on the electronic system as described above. This back-reaction ultimately also effects the forces on the nuclei again, however, such higher order effects are neglected in the present work. This formalism (Eq.~\ref{gradv}) is exact in the small displacement limit and is suited to methods where the basis itself depends upon the position of the nucleus, as in the present case where state-of-the-art highly accurate full-potential linearized augmented plane wave basis\cite{singh} is used.

Computational parameters: All calculations are performed for L1$_0$ structure of FePt\cite{lat} with lattice parameter a=3.88$\si{angstom}$ and c=3.73 $\si{angstrom}$. A {\bf k}-point set of $12\times12\times8$ was used for real-time TDDFT calculations. The phonon spectra and calculation of electron-phonon coupling requires a finer {\bf k}-point grid and we have used a grid of $24\times24\times20$. All states up to 50~eV above the Fermi energy were included in our calculations. For time propagation we have used a time step of 2.42 atto-seconds, more details of the time propagation algorithm can be found in \cite{dewhurst2016}. A smearing width of 0.027~eV was employed for the ground-state as well as for time propagation. In all calculations we use the adiabatic LDA functional, which has been shown to be highly accurate in treating very early time spin dynamics\cite{dewhurst2018,Siegrist2019,hofherr2020,clemens2020,krieger2015}. All calculations are performed using the Elk code\cite{elk}.

\section{Supplementary  information}

\paragraph*{Density of states}
\begin{figure}[ht]
\includegraphics[width=\columnwidth, clip]{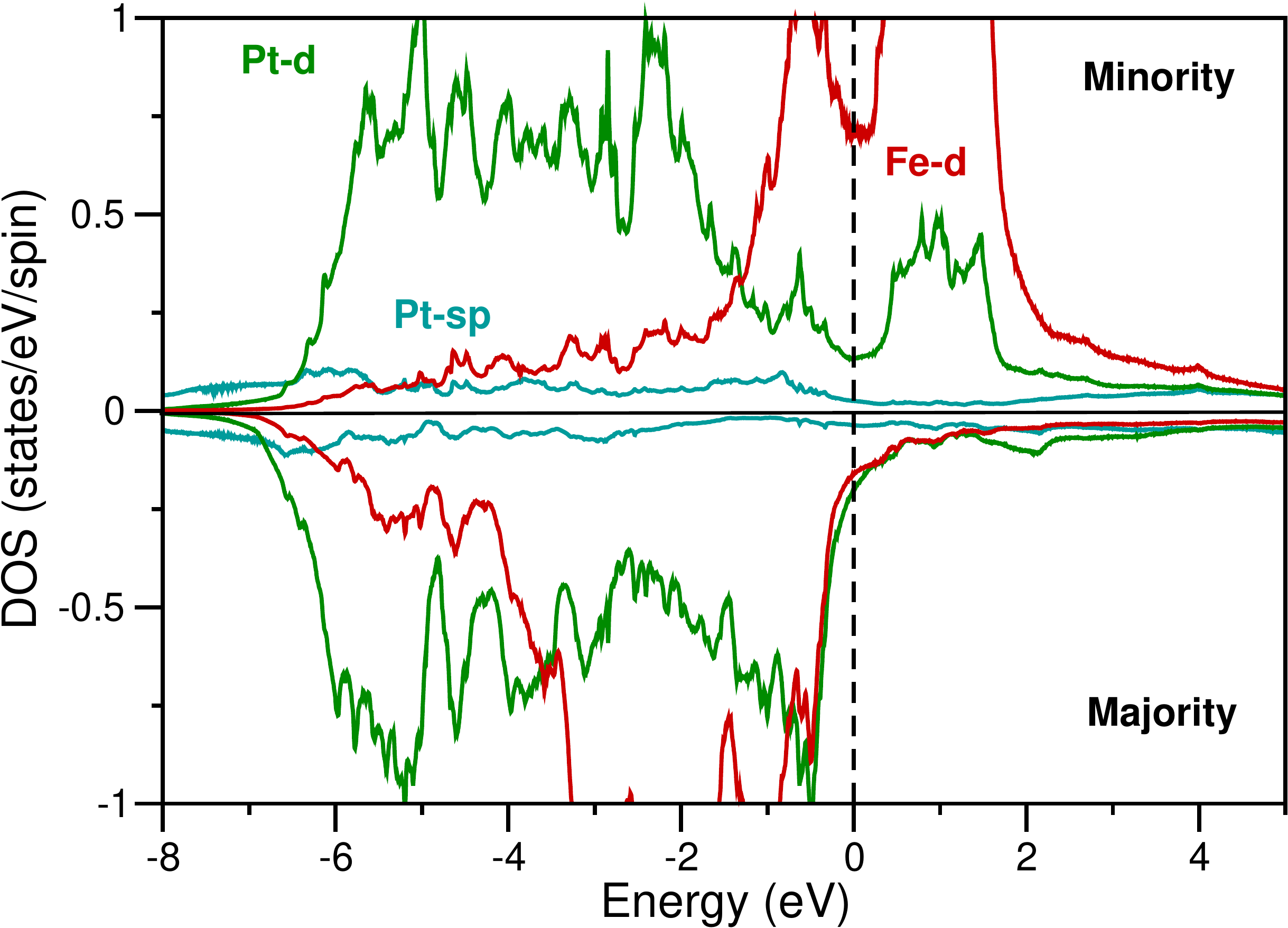}
\caption{\raggedright \raggedright Atom and spin projected ground-state density of states (in states/spin/eV) for FePt as a function of energy (in eV).} \label{dos}
\end{figure}
In Fig. \ref{dos} is shown the atom, spin and state projected ground-state density of state (DOS) for FePt. In the minority spin channel there is a large number of occupied Fe $d$-states just below and unoccupied Pt $d$-states just above the Fermi level (at 0eV). This configuration allows for minority spin current when pumped with a laser pulse of 1.55eV (800nm).

\paragraph*{Various contributions to demagnetization}
\begin{figure}[ht]
\includegraphics[width=\columnwidth, clip]{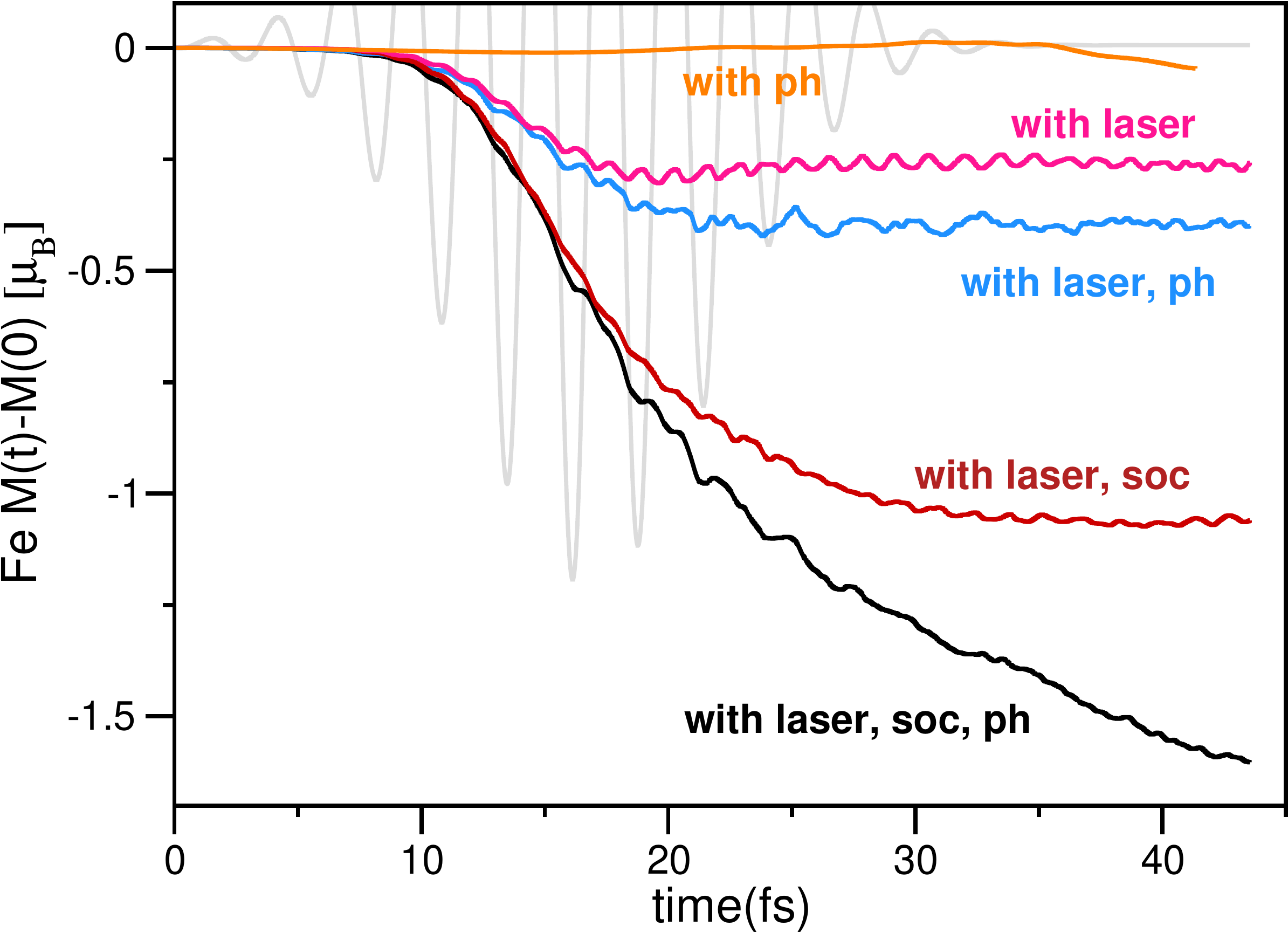}
\caption{\raggedright \raggedright Transient magnetic moment (in $\mu_B$). Calculations are performed with and without phonons, as well as with and without SOC. } \label{mom-contr}
\end{figure}
By turning off various terms in the Hamiltonian (Eq. 1 of the article) we can further analyse the effect of these terms on the spin-dynamics (see Fig.~\ref{mom-contr}). From these results it is clear that in the early times ($< 25$~fs) physics of OISTR (optical intersite spin transfer) and SOC (spin-orbit coupling) induced spin-flips dominate the mechanism of demagnetization; there is an initial increase in Pt moment and decrease in Fe moment (see Fig. 2 of the article), which happens even in the absence of SOC (see Fig. \ref{mom-contr}). However, beyond 25~fs coupled spin-nuclear dynamics starts to dominate; there is an enhanced demagnetization due to non-adiabatic coupling of nuclear and spin degrees of freedom. 

\paragraph*{Low fluence pump pulse}
\begin{figure}[ht]
\includegraphics[width=\columnwidth, clip]{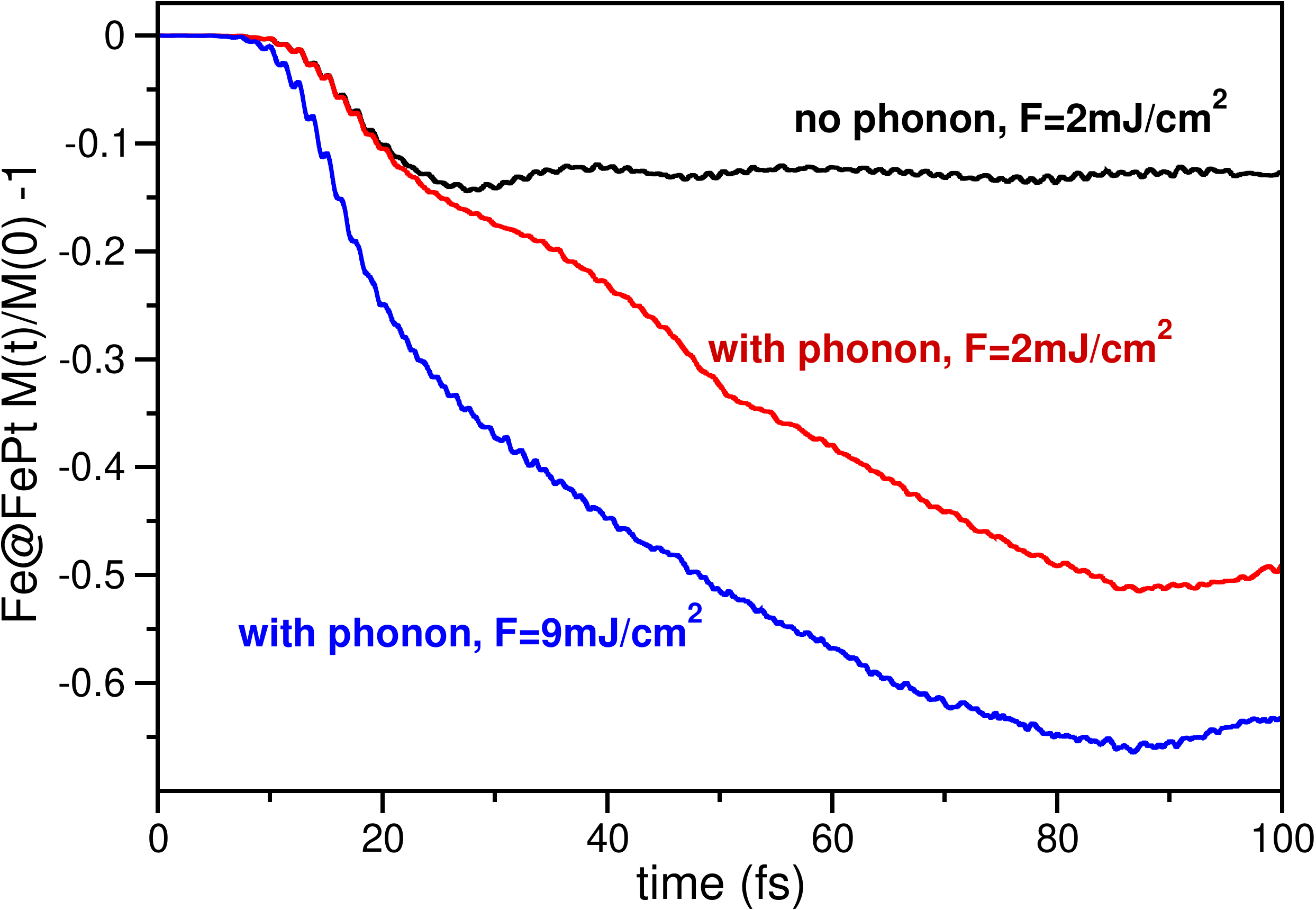}
\caption{\raggedright \raggedright Normalized atom-resolved spin moment as a function of  time (in fs) in laser pumped FePt. Two different pump pulses were used both have a central frequency of 1.55~eV and FWHM of 13~fs. One pulse has incident fluence 9~mJ/cm$^2$ and other 2~mJ/cm$^2$ . The moment is calculated in presence of the phonon modes as well as in absence of any nuclear motion. Results are shown for the $\Gamma$-point out-of-plane Fe-Pt mode with maximum amplitude of 2.4~pm. } \label{lf}
\end{figure}
Results in Fig. \ref{lf} show that the physics is also valid for low fluence pump laser pulses and is dominated by the nuclear-spin coupling with an increased demagnetization by 40\% within 40~fs

\noindent \textbf{Acknowledgements:} Sharma and JKD would like to thank DFG for funding through project-ID 328545488 TRR227 (projects A04). Shallcross would like to thank DFG for funding through SPP 1840 QUTIF Grant No. SH 498/3-1, while PE thanks DFG for finding through project 2059421. The authors acknowledge the North-German Supercomputing Alliance (HLRN) for providing HPC resources that have contributed to the research results reported in this paper.\\

%\bibliography{phmag}

%merlin.mbs apsrev4-1.bst 2010-07-25 4.21a (PWD, AO, DPC) hacked
%Control: key (0)
%Control: author (8) initials jnrlst
%Control: editor formatted (1) identically to author
%Control: production of article title (-1) disabled
%Control: page (0) single
%Control: year (1) truncated
%Control: production of eprint (0) enabled
%
\end{document}